# Quantum Computing with Variable Complex Plane. Light Beam Guide Implementation

Alexander SOIGUINE[1]

[1] SOiGUINE Quantum Computing, Aliso Viejo, CA 92656, USA

www.soiguine.com

Email address: alex@soiguine.com

**Abstract:** Following the B. Hiley belief [1] that unresolved problems of conventional quantum mechanics could be the result of a wrong mathematical structure, an alternative basic structure is suggested. Critical part of the structure is modification of the sense of commonly used terms "state", "observable", "measurement" giving them a clear unambiguous definition. This concrete definition, along with using of variable complex plane [2], is quite natural in geometric algebra terms [3]. It helps to establish a feasible language for the area of quantum computing. The suggested approach is used then in the fiber optics quantum information transferring/processing scenario.

1. Introduction

The common wisdom of conventional quantum mechanics reads something like "The particles making up our universe are inherently uncertain creatures, able to simultaneously exist in more than one place or more than one state of being." And also, certainly, "It follows from the weirdness of reality at small scales."

The weirdness has nothing to do with the scales. Wave-particle mysterious dualism follows from the lack of clear distinguishing between operators and operands.

A lot of confusion comes from the lack of precision in using terms like "state", "observable", "measurement of observable in a state", etc. This terminology creates ambiguity because the meaning of the words differs between prevailing quantum mechanics and what is logically and naturally assumed by the human mind in scientific researches and generally used in areas of physics other than quantum mechanics[1]. Nevertheless, I will try using the terminology as close as possible to commonly accepted quantum mechanics paying respects to generations of people who learned quantum mechanics in the existing framework and worked in that area of physics.

First of all, we must to strictly follow a very general definition [4] (MOS definition):

Measurement of an observable $O(\cdot)$ in a state $S(\cdot)$ is a map

$$(S(\lambda), O(\mu)) \to O(\nu)$$

---

[1] "The Copenhagen interpretation *is* quantum mechanics" (R. Peierls), see [16], page 35.

where $O(\mu)$ is an element of the set of observables; $S(\lambda)$ is an element of another set, set of states.

The sets $O(\cdot)$ and $S(\cdot)$ are not necessary different in their formal mathematical implementations. However, alignment in mathematical implementations does not mean that the sets are ontologically identical.

In general, the set of states is external to the set of observables, and vice versa, though they can be identical.

In classical computation scheme every number has a binary expansion of 0's and 1's, so we can encode any input data by bit strings. Thus with a fixed length of the strings, some $n$, we deal with vectors in $Z_2^n$. Then, in its most general form [5], classical computation can be thought of as

- The initial input $x \in Z_2^n$ encoded onto some physical system
- The evolution of $x$ processed in the physical system
- Reading out of the computational result $f(x)$ through some measurement of the system

In conventional quantum mechanics terms the three above steps of quantum computation become [6]:

- Initialize system in some known state $|\psi_0\rangle$
- Unitary evolve the system until it is in some final state $U(t)|\psi_0\rangle$
- Measure the state of the system at the end of evolution

Instead of the bit $\{0,1\} \in Z_2$ we have, in the above scheme, qubit – a quantum two-level system with two basis states $|0\rangle$ and $|1\rangle$. Qubit is formally an element of two dimensional complex Hilbert space $C^2 \ni \begin{pmatrix} z_1 \\ z_2 \end{pmatrix} = z_1 \begin{pmatrix} 1 \\ 0 \end{pmatrix} + z_2 \begin{pmatrix} 0 \\ 1 \end{pmatrix} = z_1|0\rangle + z_2|1\rangle$. The state $|\psi_0\rangle$ is then an element of $(C^2)^n$ [5], if the same assumption of a fixed length of qubit strings is made.

As was shown, see [3], [7], [8], a qubit state can be lifted to g-qubit, element of $G_3^+$ - even geometric subalgebra of the geometric algebra $G_3$ in three dimensions. The lift particularly uses the generalization of a formal imaginary plane to explicitly defined planes in three dimensions [2], [3]. The g-qubit state is strictly interpreted as operator acting on observables, also elements of geometric algebra, in the process of measurement. That follows Dirac's seminal idea [9] to remove the distinction between an element of the operator algebra and the wave function (state) without losing any information about the content of what is carried by the wave function.

Thus, the suggested computational scheme gets the form:

- Initialize system in some known state $(G_3^+)^n$, set of operators which can act on observables composed from elements of $G_3$
- Evolve the system until it is in some final state
- Identify the state of the system at the end of evolution by acting with the operators comprising the final state on observables[2]

Since the degrees of freedom of just one g-qubit give infinite number of available values, implementation of the simplest case $n=1$ would be of great importance.

In the case of electromagnetic field its state, considered as element of geometric algebra, acts (operates) on other physical entities which can also be electromagnetic fields.

## 2. Qubit states in geometric algebra

The Dirac's idea is exactly what is shown below to be an accurate implementation of the above MOS definition in the case of a g-qubit as the state in terms of geometric algebra, when the action of a state on observable is non-commutative operation

$$(S(\lambda), O(\mu)) \to O(\nu) \xleftarrow{def} O(\nu) = S^{-1}(\lambda) O(\mu) S(\lambda),$$

where $S(\lambda)$ are elements of even subalgebra $G_3^+$ of geometric (Clifford) algebra $G_3$ over three dimensional Euclidean space [3], and $O(\mu)$, $O(\nu)$ are generally elements of $G_3$.

The even subalgebra $G_3^+$, in the fiber bundle terms, can be taken as total space for base space $C^2$ and any $C^2$ qubit $\begin{pmatrix} x_1 + iy_1 \\ x_2 + iy_2 \end{pmatrix}$ has fiber in $G_3^+$. The construction is the following one.

Let $(B_1, B_2, B_3)$ is an arbitrary triple of unit value mutually orthogonal bivectors in three dimensions satisfying, with the assumption of right-hand screw orientation, the identity $B_1 B_2 B_3 = 1$[3] and multiplication rules:

$$B_1 B_2 = -B_3, B_1 B_3 = B_2, B_2 B_3 = -B_1$$

The elements of the fiber are g-qubits[4] defined as the lift:

---

[2] Due to critical reasons explained later a state should actually be a couple of a g-qubit and integer number that will eliminate ambiguity in the g-qubit angle value.

[3] The reference frame $(B_1, B_2, B_3)$ can be chosen as left-hand screw oriented, $B_1 B_2 B_3 = -1$. It is just reference frame and has nothing to do with the physical nature of three dimensional space.

[4] The element of fiber depends on which basis bivector is chosen as defining "complex plane". Cyclic permutation of the reference frame bivectors gives different elements.

$$\begin{pmatrix} x_1 + iy_1 \\ x_2 + iy_2 \end{pmatrix} \Rightarrow$$

$$x_1 + y_1 B_1 + y_2 B_2 + x_2 B_3 = x_1 + y_1 B_1 + y_2 B_1 B_3 + x_2 B_3 = x_1 + y_1 B_1 + (x_2 + y_2 B_1) B_3 \quad (2.1)$$

The fiber reference frame $(B_1, B_2, B_3)$ can be arbitrary rotated in three dimensions. In that sense we have principal fiber bundle $G_3^+ \to C^2$ with the standard fiber as group of rotations which is also effectively identified by elements of $G_3^+$.

Fiber element $x_1 + y_1 B_1 + y_2 B_2 + x_2 B_3 = x_1 + y_1 B_1 + (x_2 + y_2 B_1) B_3$ is the geometric algebra sum of two items, $x_1 + y_1 B_1$ and $(x_2 + y_2 B_1) B_3$, the first is the fiber element corresponding to conventional quantum mechanical state $|0\rangle$, in usual Dirac notations, and the second one – corresponding to $|1\rangle$.

The state $x_1 + y_1 B_1$ when acting on a $G_3^+$ observable does not change the $B_1$ component of an observable and only rotates other two components of the bivector part belonging to the subspace spanned by $B_2$ and $B_3$ [3], [7].

The state $(x_2 + y_2 B_1) B_3$ structurally differs from $x_1 + y_1 B_1$ by additional factor $B_3$. The latter makes flip of the result of the action of $x_2 + y_2 B_1$ on observable over the plane $B_1$, particularly changes the sign of the $B_1$ component.

Thus the actual geometrical sense of the $G_3^+$ fiber states corresponding to conventional quantum mechanical basis states $|0\rangle$ and $|1\rangle$ is that the first one only rotates observable around an axis orthogonal to some arbitrary given plane in three dimensions, while the second one additionally flips the result, after rotation, over that plane.

To make the $G_3^+$ notations as similar as possible to the Dirac's bra-ket ones I can also write:

$$s(\alpha, \beta, I_S) = \alpha + \beta_1 B_1 + \beta_2 B_2 + \beta_3 B_3 =$$

$$\alpha + \sqrt{\beta_1^2 + \beta_2^2 + \beta_3^2} \left( \frac{\beta_1}{\sqrt{\beta_1^2 + \beta_2^2 + \beta_3^2}} B_1 + \frac{\beta_2}{\sqrt{\beta_1^2 + \beta_2^2 + \beta_3^2}} B_2 + \frac{\beta_3}{\sqrt{\beta_1^2 + \beta_2^2 + \beta_3^2}} B_3 \right) =$$

$$\alpha + \beta(b_1 B_1 + b_2 B_2 + b_3 B_3) = \alpha + I_S \beta \equiv |s\rangle_{(\alpha, \beta, I_S)}$$

where $I_S = b_1 B_1 + b_2 B_2 + b_3 B_3$, $b_i = \frac{\beta_i}{\beta}$, $\beta = \sqrt{\beta_1^2 + \beta_2^2 + \beta_3^2}$, $\alpha^2 + \beta^2 = 1$ and $(B_1, B_2, B_3)$ - some bivector basis satisfying the above orientation and multiplication requirements. Then the conjugate is

$$\overline{\alpha + \beta_1 B_1 + \beta_2 B_2 + \beta_3 B_3} = \alpha - \beta_1 B_1 - \beta_2 B_2 - \beta_3 B_3 \equiv {}_{(\alpha, \beta, I_S)}\langle s|$$

## 3. Evolution of the g-qubit states

It is plausible to retrieve how the Hamiltonian action on states in conventional quantum mechanics is generalized in the current context.

Any conventional quantum mechanics (CQM) $C^2$ state lift in $G_3^+$ can be written as exponent:

$$\begin{pmatrix} x_1 + iy_1 \\ x_2 + iy_2 \end{pmatrix} \Rightarrow x_1 + y_1 B_1 + y_2 B_2 + x_2 B_3 = x_1 + \sqrt{1-x_1^2} \left( \frac{y_1}{\sqrt{1-x_1^2}} B_1 + \frac{y_2}{\sqrt{1-x_1^2}} B_2 + \frac{x_2}{\sqrt{1-x_1^2}} B_3 \right) =$$

$$\cos \varphi + (b_1 B_1 + b_2 B_2 + b_3 B_3) \sin \varphi = e^{I_S \varphi}$$

where $\cos \varphi = x_1$, $\sin \varphi = \sqrt{1-x_1^2}$, $b_1 = \frac{y_1}{\sqrt{1-x_1^2}}$, $b_2 = \frac{y_2}{\sqrt{1-x_1^2}}$, $b_3 = \frac{x_2}{\sqrt{1-x_1^2}}$,

$$I_S = b_1 B_1 + b_2 B_2 + b_3 B_3.$$

Hamiltonian in CQM is a self adjoint matrix of general form:

$$H = \begin{pmatrix} \alpha + \beta_1 & \beta_2 - i\beta_3 \\ \beta_2 + i\beta_3 & \alpha - \beta_1 \end{pmatrix}$$

It acts on two dimensional complex vectors by usual rules of linear algebra, $H|\psi\rangle$.

The Hamiltonian matrix geometric algebra lift[5] is not element of $G_3^+$, since it has the form $\alpha + I_3(\beta_1 B_1 + \beta_2 B_2 + \beta_3 B_3)$. We can forget about $\alpha$ because it only may cause final multiplication by a scalar. Then

$$H_G \stackrel{def}{\equiv} I_3(\beta_1 B_1 + \beta_2 B_2 + \beta_3 B_3) =$$

$$I_3 \sqrt{\beta_1^2 + \beta_2^2 + \beta_3^2} \left( \frac{\beta_1}{\sqrt{\beta_1^2 + \beta_2^2 + \beta_3^2}} B_1 + \frac{\beta_2}{\sqrt{\beta_1^2 + \beta_2^2 + \beta_3^2}} B_2 + \frac{\beta_3}{\sqrt{\beta_1^2 + \beta_2^2 + \beta_3^2}} B_3 \right) =$$

$$I_3 |H_G| I_{H_G} = I_3 |H_G| e^{\frac{\pi}{2} I_{H_G}}$$

Thus we see that multiplication of a complex two dimensional vector by matrix $H$ corresponds, if mapped directly to multiplication in $G_3$, to the operation:

$$I_3 |H_G| e^{\frac{\pi}{2} I_{H_G}} e^{I_S \varphi}$$

---

[5] See the Hamiltonian lift calculation in [3]

It looks not good since the result does not belong to $G_3^+$, our space of states. It means that the action of Hamiltonians, as matrices, on the $C^2$ states as linear algebra multiplication, cannot be equivalent to multiplication of the results of the lifts in $G_3$.

We have two options of lifting the operation $H\begin{pmatrix} x_1 + iy_1 \\ x_2 + iy_2 \end{pmatrix}$ to $G_3^+$:

- Rotation of a $G_3^+$ element, particularly a state, in the plane of the Hamiltonian lift by the angle defined by the Hamiltonian value.
- Clifford translation of a $G_3^+$ element, particularly a state, along big circle of the $S^3$ sphere. The circle is intersection of the sphere with the plane of the Hamiltonian lift.

Let's initially consider the second option.

Instead of unitary transformations acting on the Hilbert space vector states of $C^2$ transforming them into new states, $U(t)|\psi_0\rangle$, the corresponding transformations acting in the fiber bundle with total space $G_3^+$ over base space $C^2$ are given, if the Hamiltonian depends on time, as sequences of infinitesimal Clifford translations [3][6]:

$$|s(t+\Delta t)\rangle_{(\alpha,\beta,I_S)} = e^{-I_3 H(t)\Delta t}|s(t)\rangle_{(\alpha,\beta,I_S)} = e^{-\left(I_3 \frac{H(t)}{|H(t)|}\right)|H(t)|\Delta t}|s(t)\rangle_{(\alpha,\beta,I_S)}$$

where $I_3$ is unit value oriented volume in the three dimensions and $H(t)$ - the Hamiltonian expanded in basis $(I_3 B_1, I_3 B_2, I_3 B_3)$. Unit value bivector $I_3 \frac{H(t)}{|H(t)|}$ is generalization of imaginary unit explicitly defining the plane of $S^3$ sphere big circle.

*Remark 3.1*: If the Hamiltonian does not depend on time a finite Clifford translation gives:

$$|s(t)\rangle_{(\alpha,\beta,I_S)} = e^{-I_3 Ht}|s(0)\rangle_{(\alpha,\beta,I_S)} = e^{-\left(I_3 \frac{H}{|H|}\right)|H|t}|s(0)\rangle_{(\alpha,\beta,I_S)}$$

The geometric algebra framework with an arbitrary variable plane of state bivector (VPSB) generalizes geometrically unspecified complex plane of CQM. Thus, it follows that the CQM Schrodinger equation $\hat{H}|\psi(t)\rangle = i\frac{\partial}{\partial t}|\psi(t)\rangle$ in the VPSB framework takes the form:

$$I_3 H(t)|s(t)\rangle_{(\alpha,\beta,I_S)} = -\frac{\partial}{\partial t}|s(t)\rangle_{(\alpha,\beta,I_S)}$$

---

[6] I will use in the following $H$ instead of $H_G$ since only the geometric algebra meaning of a Hamiltonian will be used without making any confusion

with generally varying bivector $I_3 H(t)$. It follows that arbitrary state transformation is the holonomy $\int_L e^{I_{H(l)}|H(l)|dl}|s\rangle_{(\alpha,\beta,I_S)}$ where the integral is taken along the Hamiltonian vector curve trace on the surface of unit sphere $S^3$ [3].

The critical thing to remember: Schrodinger equation in geometric algebra terms is an equation defining evolution of states, operators. The states act on other states either via the Clifford translations or on states interpreted as observables when executing measurements (see the above MOS definition).

The option of rotation in the plane of the Hamiltonian $G_3^+$ lift by the angle defined by Hamiltonian value will be considered after the next section on electric field polarization.

## 4. Electric field polarization

To deal with the guided light beams as physical processes carrying information about the states in the geometric algebra sense I shall begin with the electromagnetic fields and their polarizations in the $G_3^+$ terms [10], [11].

What is different in the current approach to the light propagation in a beam guide is the fact that formally used imaginary unit is replaced with a unit bivector in three dimensional space not necessary orthogonal to the $z$ direction, default beam guide axis.[7] The electric fields should naturally be considered as states, up to the magnitude factor, that's the $G_3^+$ operators acting on observables.

Assume we deal with a detectable polarization in the $xy$ plane: circular, elliptic or linear one, which means that the electric field vector end point moves along the corresponding trajectory. The following result takes place:

Theorem 4.1. Any type of polarization in the $xy$ plane is projection of circular polarization in some plane $S$.

*Proof:*

Since the plane of rotation/oscillation of electric field vector may be any plane in three dimensions, the plane of polarization should be explicitly defined[8].

Electric field can evolve in a plane of some unit bivector $I_S$ being in the state of circular polarization in that plane.

Suppose a polarization is measured in $xy$ plane and is an ellipse of general parametric form:

$$(a\cos\alpha\cos t - b\sin\alpha\sin t)\hat{x} + (a\sin\alpha\cos t + b\cos\alpha\sin t)\hat{y}, \quad 0 \le t \le 2\pi, \quad (4.1)$$

---

[7] The interest to the transverse light beam spin models is growing intensively, see, for example [15]
[8] Similar definitions of polarization are used in some different contents, see, for example [14].

where $\alpha$ is angle of the ellipse $a\cos t\hat{x} + b\sin t\hat{y}$, $0 \leq t \leq 2\pi$, rotation in $xy$ plane relative to the $x$ direction, $a$ is value of the ellipse semi axis along the direction of the $x$ axis, $b$ is semi axis along the orthogonal direction, $\hat{x}$ and $\hat{y}$ are unit vectors along corresponding axes.

*Remark 4.1:* In pure geometric algebra terms the rotation of ellipse with semi axes parallel to coordinate axes by angle $\alpha$ is $(a\cos t\hat{x} + b\sin t\hat{y})e^{\hat{x}\hat{y}\alpha}$ (multiplication from the right!).

*Remark 4.2:* If $a = b$ (circle) the rotation gives the same circle. If one of semi axes, say $b$, is zero, we get vector oscillating with the amplitude $a$ along the line $(\cos\alpha)\hat{x} + (\sin\alpha)\hat{y}$ (or with the amplitude $b$ along the line $-(\sin\alpha)\hat{x} + (\cos\alpha)\hat{y}$ if $a = 0$). This is the case of a linear polarization.

Assume the normal to $I_S$ be received from the normal to $xy$ plane ($z$ direction) by rotation by angle $\theta$ in a plane $S_R$ passing through the major semi axis of ellipse (4.1). Define angle of rotation by $\cos(\theta) = \dfrac{\min(a,b)}{\max(a,b)}$. The plane of rotation is defined by unit bivector dual to unit vector along minor semi axis. If major semi axes is of value $a$ then

$$I_{S_R} = -(\sin\alpha)I_3\hat{x} + (\cos\alpha)I_3\hat{y}$$

If major semi axes has value $b$,

$$I_{S_R} = (\cos\alpha)I_3\hat{x} + (\sin\alpha)I_3\hat{y}$$

Thus, the two unit bivectors for the $xy$ plane and $S$ plane are received from each other as:

$$I_S = e^{-I_{S_R}\frac{\theta}{2}} I_{xy} e^{I_{S_R}\frac{\theta}{2}}, \quad I_{xy} = e^{I_{S_R}\frac{\theta}{2}} I_S e^{-I_{S_R}\frac{\theta}{2}}$$

The projection of the $I_S$ polarization circle, expanded to radius $\max(a,b)$, onto $xy$ plane is exactly the original ellipse (4.1).

*QED.*

The circular polarized electromagnetic wave states actually comprise the basis for all other types of polarizations because they are the only type of waves following from the solution of Maxwell equations in free space accurately done in geometric algebra terms.

Let's take the electromagnetic field in the form

$$F = F_0 \exp\left[I_S\left(\omega t - \vec{k}\cdot\vec{r}\right)\right] \qquad (4.2)$$

with the only requirement that it satisfies the Maxwell system of equations in free space, which in geometrical algebra terms takes the form of one equation:

$$(\partial_t + \nabla)F = 0$$

where $\nabla = \frac{\partial}{\partial x}\hat{x} + \frac{\partial}{\partial y}\hat{y} + \frac{\partial}{\partial z}\hat{z}$ and the multiplication is the geometrical algebra one.

Element $F_0$ in (4.2) is a constant element of geometric algebra $G_3$, undefined yet, and $I_S$ is unit value bivector of a plane $S$ in three dimensions, generalization of the imaginary unit in the current approach.

In geometric algebra terms electromagnetic field can be identified by geometric algebra sum of a vector $E$, the electric field, and bivector $I_3 B$, magnetic field. That means that to retrieve the structure of the element $F_0$ we need to compare the right hand side of (4.1) with the geometric algebra element $E + I_3 B$.

The exponent in (4.2) is unit value element of $G_3^+$ with the $I_S$ bivector plane, that's $e^{I_S \varphi} = \cos(\varphi) + I_S \sin(\varphi)$, $\varphi = \omega t - \vec{k} \cdot \vec{r}$. Since no assumptions about $F_0$ have been done yet we will generally write: $F_0 = (F_0)_s + (F_0)_V + (F_0)_B + (F_0)_P$.

The geometric algebra product $F_0 e^{I_S \varphi} = [(F_0)_s + (F_0)_V + (F_0)_B + (F_0)_P][\cos(\varphi) + I_S \sin(\varphi)]$ should give $E + I_3 B$ which is sum of a vector and bivector.

Firstly it follows that $(F_0)_s$ and $(F_0)_P$ must be zeros.

Secondly, the product $(F_0)_V I_S$ is sum of a vector and pseudoscalar (see [3], Sec.1.3). The pseudoscalar must be zero that is only possible when $(F_0)_V$ lies in the plane of $I_S$. The remaining vector part of the product is equal to the cross product of $(F_0)_V$ and vector dual to $I_S$, that's $(F_0)_V \wedge I_3 I_S$ which is the vector $(F_0)_V$ rotated by $\frac{\pi}{2}$ in the positive direction in plane $S$.

Thirdly, the product of two bivectors $(F_0)_B I_S$ is sum of the scalar, equal to scalar product of two vectors dual correspondingly to $(F_0)_B$ and $I_S$, and bivector dual to vector $I_3 (F_0)_B \wedge I_3 I_S$. The scalar part must be zero which means that the bivector planes are orthogonal. Then the remaining bivector $I_3 (I_3 (F_0)_B \wedge I_3 I_S)$ is the $(F_0)_B$ rotated by $\frac{\pi}{2}$ around the axis orthogonal to the plane $S$.

Thus, the geometric algebra element $F$ is geometric algebra sum of a vector in plane $S$ and bivector orthogonal to that plane. Both rotate synchronically with the angle $\varphi = \omega t - \vec{k} \cdot \vec{r}$ around axis orthogonal to plane $S$ and lying in the plane of the bivector.

It is interesting to reveal what further follows from the requirement $(\partial_t + \nabla) F = 0$, Maxwell equation.

The derivative by time gives

$$\frac{\partial}{\partial t}F = F_0 e^{I_S\varphi}I_S\frac{\partial}{\partial t}\left(\omega t - \vec{k}\cdot\vec{r}\right) = F_0 e^{I_S\varphi}I_S\omega = FI_S\omega.$$

The geometric algebra product $\nabla F$ is:

$$\nabla F = F_0 e^{I_S\varphi}I_S\nabla\left(\omega t - \vec{k}\cdot\vec{r}\right) = -F_0 e^{I_S\varphi}I_S\vec{k} = -FI_S\vec{k},$$

and the Maxwell equation becomes:

$$F\left(I_S\omega - I_S\vec{k}\right) = 0$$

or:

$$F\left(I_S\omega + \vec{k}I_S\right) = 0$$

The geometrical product $\vec{k}I_S$ is the sum of the $\vec{k}$ component, vector, in plane $S$ rotated by $\frac{\pi}{2}$, and pseudoscalar of volume equal to the length of the $\vec{k}$ component orthogonal to plane $S$. I will denote the vector item as $\vec{k}_S\left(\frac{\pi}{2}\right)$ and the pseudoscalar as $I_3|\vec{k}_\perp|$.

The field $F$ should be the sum of vector field and bivector field, $E + I_3 B$. The geometric product $EI_S\omega$ is again the sum of vector $E_S\left(\frac{\pi}{2}\right)\omega$ and pseudoscalar $I_3|E_\perp|\omega$. In the similar way, $I_3 BI_S\omega$ is the sum of bivector dual to vector $B_S\left(\frac{\pi}{2}\right)\omega$ and scalar $-\omega|B_\perp|$.

The products of $F$ with the $\vec{k}I_S$ are comprised of the following. $E\vec{k}I_S$ is sum of scalar $E\cdot\vec{k}_S\left(\frac{\pi}{2}\right)$, bivector $I_3\left(E\times\vec{k}_S\left(\frac{\pi}{2}\right)\right)$ and bivector $I_3 E|\vec{k}_\perp|$. Similarly, $I_3 B\vec{k}I_S$ is sum of $I_3\left(B\cdot\vec{k}_S\left(\frac{\pi}{2}\right)\right)$, $-B\times\vec{k}_S\left(\frac{\pi}{2}\right)$ and $-B|\vec{k}_\perp|$. Combining all that we have in the left-hand side of the Maxwell equation the sum of:

scalar: $\quad -\omega|B_\perp| + E\cdot\vec{k}_S\left(\frac{\pi}{2}\right)$

vector: $\quad \omega E_S\left(\frac{\pi}{2}\right) - B\times\vec{k}_S\left(\frac{\pi}{2}\right) - B|\vec{k}_\perp|$

bivector: $\quad \omega I_3 B_S\left(\frac{\pi}{2}\right) + I_3 E\times\vec{k}_S\left(\frac{\pi}{2}\right) + I_3 E|\vec{k}_\perp|$

pseudoscalar: $\quad I_3\omega|E_\perp| + I_3\left(B\cdot\vec{k}_S\left(\frac{\pi}{2}\right)\right)$

Thus we get four equations for the four components of the geometric algebra element $F$ of $G_3$:

$$-\omega|B_\perp| + E \cdot \vec{k}_S\left(\frac{\pi}{2}\right) = 0$$

$$\omega|E_\perp| + B \cdot \vec{k}_S\left(\frac{\pi}{2}\right) = 0$$

$$\omega E_S\left(\frac{\pi}{2}\right) - B \times \vec{k}_S\left(\frac{\pi}{2}\right) - B|\vec{k}_\perp| = 0$$

$$\omega B_S\left(\frac{\pi}{2}\right) + E \times \vec{k}_S\left(\frac{\pi}{2}\right) + E|\vec{k}_\perp| = 0$$

The above equations can be used as additional conditions on vectors $E$ and $B$ (when the vector $\vec{k}$ is assumed to be given) to the earlier received statement that, particularly, vector $E$ rotates in plane $S$ with the varying angle $\varphi = \omega t - \vec{k} \cdot \vec{r}$. This rotation defines circular polarization in plane $S$, thus justifying practical applicability of the earlier results that any polarization in the $xy$ plane can be received as projection of circular polarization in some plane.

## 5. Hamiltonian action as rotation

An electric field defined by vector rotating in a plane $S$ is obviously a state (up to real constant multiplier, amplitude) in the $G_3^+$ terms.

Below we consider the situation that is usually defined as spin and orbital angular momenta, or chirality in other terminology.

An arbitrary spin angular momentum is defined by the result of inclination of electric field vector rotating in the $xy$ plane. The orbital angular momentum appears when the inclined plane rotates around the $z$ axis. Thus we have composition of inclination of unit bivector, $I_S = e^{-I_{S_R}\frac{\theta}{2}} I_{xy} e^{I_{S_R}\frac{\theta}{2}}$, and further rotation.

The plane of initial inclination $I_{S_R} = -(\sin\alpha)I_3\hat{x} + (\cos\alpha)I_3\hat{y}$ can be taken with $\alpha = 0$ because the projection polarization ellipse rotates, by default permanently, in the $xy$ plane.

So we assume $I_{S_R}\big|_{\alpha=0} = -(\sin\alpha)I_3\hat{x} + (\cos\alpha)I_3\hat{y}\big|_{\alpha=0} = I_3\hat{y} = I_{zx}$. With the usual identification of basis bivectors: $B_1 = \hat{y}\hat{z}$, $B_2 = \hat{z}\hat{x}$, $B_3 = \hat{x}\hat{y}$, we get[9] $I_S = e^{-B_2\frac{\theta}{2}} B_3 e^{B_2\frac{\theta}{2}}$:

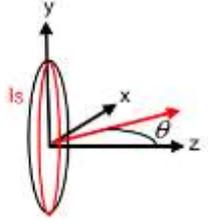

Rotation of the inclined polarization state around the $z$ axis gives the result:

$$e^{-B_3\frac{\varphi(t)}{2}} e^{-B_2\frac{\theta}{2}} B_3 e^{B_2\frac{\theta}{2}} e^{B_3\frac{\varphi(t)}{2}}$$

where $\varphi(t)$ is varying with time angle of rotation around the $z$ axis.

The inclination gives $B_1 \sin\theta + B_3 \cos\theta$. Then subsequent result of the state rotation around the $z$ axis has explicit bivector form:

$$U(\theta,\varphi) = B_1 \sin\theta \cos\varphi + B_2 \sin\theta \sin\varphi + B_3 \cos\theta = B_1 e^{B_3\varphi} \sin\theta + B_3 \cos\theta$$

The derivatives $\dfrac{\partial U}{\partial \theta} = B_1 e^{B_3\varphi} \cos\theta - B_3 \sin\theta$ and $\dfrac{\partial U}{\partial \varphi} = B_2 e^{B_3\varphi} \sin\theta$ give two Berry connections:

$$A_\theta = \overline{U} B_3 \frac{\partial U}{\partial \theta} = -\left(e^{-B_3\varphi} B_1 \sin\theta + B_3 \cos\theta\right) B_3 \left(B_1 e^{B_3\varphi} \cos\theta - B_3 \sin\theta\right) =$$

$$-\left(e^{-B_3\varphi} B_2 \sin\theta - \cos\theta\right)\left(B_1 e^{B_3\varphi} \cos\theta - B_3 \sin\theta\right) =$$

$$-e^{-B_3\varphi} B_3 e^{B_3\varphi} \sin\theta\cos\theta + B_1 e^{B_3\varphi} \cos^2\theta - e^{-B_3\varphi} B_1 \sin^2\theta - B_3 \sin\theta\cos\theta =$$

$B_1 e^{B_3\varphi} \cos 2\theta - B_3 \sin 2\theta$ [10],

$$A_\varphi = \overline{U} B_3 \frac{\partial U}{\partial \varphi} = -\left(e^{-B_3\varphi} B_1 \sin\theta + B_3 \cos\theta\right) B_3 B_2 e^{B_3\varphi} \sin\theta =$$

$$\sin^2\theta + B_2 e^{B_3\varphi} \sin\theta\cos\theta = \frac{1}{2}\left(1 - \cos 2\theta + B_2 e^{B_3\varphi} \sin 2\theta\right)$$

---

[9] We ignore for convenience that the inclined bivector has not unit value and more accurately should be $|E| e^{-B_2\frac{\theta}{2}} B_3 e^{B_2\frac{\theta}{2}}$

[10] Easy to verify: $B_1 e^{B_3\varphi} = e^{-B_3\varphi} B_1$, that was used in calculations

with the corresponding Berry curvature:

$$F_{\theta\varphi} = \partial_\theta A_\varphi - \partial_\varphi A_\theta = \sin 2\theta$$

The connections differ from the CQM case by additional bivector terms since they were calculated in more detailed formalism, though the curvature naturally remains identical because it only depends on the topology.

One of the ideas of practical implementation of topological quantum computations is using of stable values of Berry connections. Let's take, as an example, one of the calculated above:

$$A_\varphi = \frac{1}{2}\left(1 - \cos 2\theta + B_2 e^{B_3\varphi} \sin 2\theta\right)$$

and consider the case $\theta = \frac{\pi}{2}$ corresponding to linear polarization rotating in the $xy$ plane. The connection value is $A_\varphi\big|_{\theta=\frac{\pi}{2}} = 1$.

Since we are in the frame of the $G_3^+$ paradigm, the only variables which can be used as keeping stable quantum topological results should be the results of measurements of observables in given states. In the considered case

$$A_\varphi\big|_{\theta=\frac{\pi}{2}} = \left(\overline{U}B_3 \frac{\partial U}{\partial \varphi}\right)\bigg|_{\theta=\frac{\pi}{2}} = -e^{-B_3\varphi} B_1 B_3 B_2 e^{B_3\varphi} = B_1 e^{B_3\varphi} B_2 B_3 e^{B_3\varphi} = B_1 B_2 e^{-B_3\varphi} B_3 e^{B_3\varphi},$$

that is, up to the factor $B_1 B_2$, the result of measurement of observable $B_3$ in the state $e^{B_3\varphi}$.

6. Rotation of the circular polarization plane and linear polarization in the projection plane

Consider again the case $\theta = \frac{\pi}{2}$ which results in linear polarization along the line rotated by $\varphi$ angle relative to $x$ axis in the $xy$ plane. The circular polarization plane contains the $z$ axis and is rotated around it with the second (external) rotation in the transformation, measurement of $B_3$ in the state $e^{B_2\frac{\pi}{4}} e^{B_3\frac{\varphi}{2}}$:

$$e^{-B_3\frac{\varphi}{2}}\left(e^{-B_2\frac{\pi}{4}} B_3 e^{B_2\frac{\pi}{4}}\right) e^{B_3\frac{\varphi}{2}}$$

Assume we have physical mechanism of rotating circular polarized electric field, state, in the $xy$ plane, in other words a mechanism sufficient to executing the measurements:

$$e^{-B_3\frac{\varphi}{2}}\left(e^{-B_2\frac{\pi}{4}}B_3 e^{B_2\frac{\pi}{4}}\right)e^{B_3\frac{\varphi}{2}} = e^{-B_3\frac{\varphi}{2}}B_1 e^{B_3\frac{\varphi}{2}} = B_1\cos\varphi + B_2\sin\varphi \qquad (6.1)$$

The result has zero value bivector component in the plane of $B_3$, as it should be, though geometrically the projection of the circle onto $xy$ plane is a degenerated ellipse - straight segment of unit length centered at 0 of the line along the vector $(\cos\varphi)\hat{x} + (\sin\varphi)\hat{y}$. Obviously, this linear polarization line rotates together with angle $\varphi$. Information about circular polarization sense is lost.

Generally we have two circular polarization states, left-hand and right-hand, and the above formula for the opposite circular polarization sense is:

$$e^{-B_3\frac{\varphi}{2}}\left(e^{-B_2\frac{\pi}{4}}(-B_3)e^{B_2\frac{\pi}{4}}\right)e^{B_3\frac{\varphi}{2}} = e^{-B_3\frac{\varphi}{2}}(-B_1)e^{B_3\frac{\varphi}{2}} = -B_1\cos\varphi - B_2\sin\varphi \qquad (6.2)$$

Thus, the line of linear polarization in the $xy$ plane remains the same. This fact is a separate problem because distinguishing between the two circular polarizations which both can be the origin of the same linear polarization is critically important for a basic algorithm of function value calculations that will be demonstrated in following section. Thus we are making the assumption that there is only one circularly polarized mode of the spin angular momentum, say with the $B_1$ sense.

## 7. Evolving states via transformations of circular polarization states

Now assume that transformation of the state $B_1$, spin angular momentum, is made by the Clifford translation with a Hamiltonian $H(t)$ as formulated earlier:

$$|s(t+\Delta t)\rangle_{(\alpha,\beta,I_S)} = e^{-I_3 H(t)\Delta t}|s(t)\rangle_{(\alpha,\beta,I_S)} = e^{-\left(I_3\frac{H(t)}{|H(t)|}\right)|H(t)|\Delta t}|s(t)\rangle_{(\alpha,\beta,I_S)}$$

To keep up with the orbital angular momenta corresponding the external transformation of the polarization plane the plane of bivector $I_3\frac{H(t)}{|H(t)|}$ is supposed to be constant and equal to $B_3$. Physically, such Hamiltonian action can be implemented via magnetic field parallel to the light guide, for example as an electric current coil around the light guide.

The core of quantum computing should not be in entanglement, which only formally follows in conventional quantum mechanics from representation of the many particle states as tensor products of individual particle states. The core of quantum computing scheme should be in manipulation and transferring of quantum states as operators decomposed in geometrical algebra variant of qubits (g-qubits), or four dimensional unit sphere points, if you prefer. In this way quantum computer is, as it should be, an analog computer keeping information in sets of objects with infinite number of degrees of freedom, contrary to the two value bits or two dimensional Hilbert space elements, qubits.

In the suggested computational scheme, defined in Sec.1, we write the initial state $\left(G_3^+\right)^n$ as $\left(e^{I_{S_1}\varphi_1}, e^{I_{S_2}\varphi_2}, ..., e^{I_{S_n}\varphi_n}\right)$. In the same way as traditional Turing machine scheme is pictured, one can schematically represent a $\left(G_3^+\right)^n$ state as

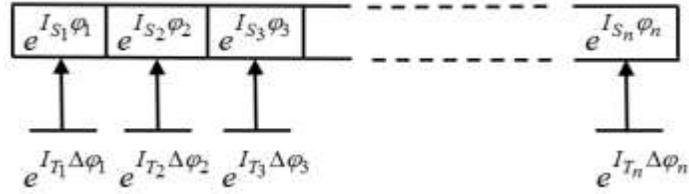

States $e^{I_{T_k}\Delta\varphi_k}$ realizing evolution, act on the components $e^{I_{S_k}\varphi_k}$ of initial state as Clifford translations:

$$e^{I_{S_k}\varphi_k} \to e^{I_{T_k}\Delta\varphi_k} e^{I_{S_k}\varphi_k} \qquad (7.1)$$

If a continuous sequence of such translations takes place we get the holonomy formulated in Sec.3:

$$\int_L e^{I_{T(l)}d\varphi(l)} |s\rangle_{(\cos\varphi(l),\sin\varphi(l),I_{S(l)})}.$$

If the transformation (7.1) is taken as an infinitesimal one (or with not varying plane $T_k$), the state $e^{I_{S_k}\varphi_k}$ is rotated in the plane $T_k$ by the angle $\Delta\varphi_k$ and synchronically rotated by the same angle in two planes orthogonal to $T_k$ in three dimensions [3].

Suppose we have a light guide with the input of series of the length $n$ time bins bearing the states $e^{I_{S_k}\varphi_k}$ [11]. The time bin items are transformed by the rules (7.1). The output state, final state in the terms of suggested scheme, acts on $n$ copies of the observable $B_1$. The result is series of the length $n$ of linear polarizations in the $xy$ plane.

All that is true in the simple considered case of the spin angular momentum orthogonal to the $z$ axis. Other arbitrary directions will give more sophisticated scenes and options.

The suggested computational scheme is applicable, for example, to function calculations.

The light guide single mode input, discrete in time, of the calculated function argument is identified by the time step number (index, time stamp $t_k$) and the time bin state item plane

---

[11] Similar time bins scheme, though with much simpler bin items, was considered, for example, in [12]

and angle. In the current case the latter two are $B_3$ and angle of rotation of the polarization $B_1$ around the $z$ axis.

Clifford translations acting on the state items all have the same plane, $T_k = B_3$, and angles of rotations are defined by Hamiltonian values $|H(t_k)|$. In the output we have a sequence of length $n$ of the final state items $e^{I_{B_3}|H(t_k)|}e^{I_{B_3}t_k}$, $1 \leq k \leq n$.

The measurement phase (the last item of our computational scheme defined in Sec.1) is the set of measurements:

$$e^{-I_{B_3}(t_k+|H(t_k)|)}B_1 e^{I_{B_3}(t_k+|H(t_k)|)}$$

giving the sequence of length $n$ of linear polarizations in the $xy$ plane as defined by (6.1):

$$B_1\cos[2(t_k+|H(t_k)|)]+B_2\sin[2(t_k+|H(t_k)|)], \ 1\leq k \leq n \ [12]$$

## 8. Conclusions

Two seminal ideas – variable and explicitly defined complex plane in three dimensions, and the $G_3^+$ states[13] as operators acting on observables – allow to put forth comprehensive and much more detailed formalism appropriate for quantum mechanics in general and particularly for quantum computing schemes. Based on this new mathematical structure a computational scheme was suggested, implemented in terms of the guided light polarization variant of geometric algebra g-qubits. The approach may be thought about, for example, as a far going geometric algebra generalization of some proposals for quantum computing formulated in terms of light beam time bins, see [12], [13], but giving much more strength and flexibility in practical implementation.

---

[12] Default righthandscrew circular polarization in the spin angular momentum plane is taken. The ambiguities in the values of function arguments and function values due to the fact that both have ranges in finite intervals can be removed via $t_k \to t_k(\mathrm{mod}(2\pi))$ and $|H(t_k)| \to |H(t_k)|(\mathrm{mod}(2\pi))$.

[13] Good to remember that state and wave function are actually (at least should be) synonyms in conventional quantum mechanics